Manuscript under review. This document has yet to be peer-reviewed. (preprint v1.0)

Artificial General Intelligence, Existential Risk, and Human Risk Perception

David R. Mandel

Intelligence, Influence, and Collaboration Section, Defence Research and Development Canada

Department of Psychology, York University

Correspondence: drmandel66@gmail.com

Abstract

Artificial general intelligence (AGI) does not yet exist, but given the pace of technological development in artificial intelligence, it is projected to reach human-level intelligence within roughly the next two decades. After that, many experts expect it to far surpass human intelligence and to do so rapidly. The prospect of superintelligent AGI poses an existential risk to humans because there is no reliable method for ensuring that AGI goals stay aligned with human goals. Drawing on publicly available forecaster and opinion data, the author examines how experts and non-experts perceive risk from AGI. The findings indicate that the perceived risk of a world catastrophe or extinction from AGI is greater than for other existential risks. The increase in perceived risk over the last year is also steeper for AGI than for other existential threats (e.g., nuclear war or human-caused climate change). That AGI is a pressing existential risk is something on which experts and non-experts agree, but the basis for such agreement currently remains obscure.

**Introduction**

It is estimated that over 99% of species that have lived on our planet are extinct (Jablonski, 2004). Earth has had at least five mass extinctions in the last half-billion years and yet it is estimated that these account for less than 5% of the total number of extinctions (Erwin, 2001). All mass extinctions were caused by extreme climatic conditions; i.e., with the notable exception of the recent loss of biodiversity at a rate considerably higher than the background extinction rate—a phenomenon that some scientists believe represents the early phase of a sixth mass extinction caused by anthropogenic factors (Ceballos et al., 2015; Wake & Vrendenberg, 2008). If so, it will be the first time an Earth species has triggered a mass extinction.

Extinction, it would seem, much like evolution, is the norm over geological time. For the human species to survive it must either preempt existential risks or successfully confront them before they become causes of human extinction. This realization, too, is something that humans—and humans alone—must confront. It drives ethical concern not only for our own future but also for the collective future of humanity over the long term (MacAskill, 2022; Ord, 2020). Technological advances are necessary for long-term human survival. In prescientific eras, humans could rely on folk beliefs about salvation from divine powers or simply wonder if the heavens were eternal. Today, however, it is collective knowledge in literate societies that the universe itself is aging (Dyson, 1979). Even if we could successfully prevent bolide collisions (which appear to have played a triggering role in 3/5 of the "big five" mass extinctions) and prevent or control other equally extreme climatic events, Earth has its days numbered. In roughly 5-6 billion years, without the intervention of intelligent life, the Sun will transition to a red giant enveloping several planets including our own. Humans must develop technologies that can alter the fate of stars or they must become a spacefaring species that can find new habitable environments in time to avoid extinction. Science (and astrophysics, more specifically) has removed all uncertainty about the inevitable long-term doom that awaits us unless we do something to avert it. That something must involve advanced technologies that currently do not exist.

Indeed, as the physicist, David Deutsch points out, the fate of stars, and the Sun in particular, cannot be determined by what we currently know about astrophysics because we cannot predict the effects of human or other future forms of consciousness on the evolution of physical processes, including ones that currently seem impervious to human intervention. According to Deutsch (1997), the task is not impossible and humans have several billion years to work out the details. However, Gott's (1993) Copernican estimate that our species has a total longevity between 0.2 and 8.0 million years with 95% confidence significantly reduces our completion time (and perhaps our collective optimism). As well, the timespan over which intelligent life might affect the trajectory of cosmological events is deeply uncertain because we do not know whether life is matter-bound or "merely" structure-bound (Dyson, 1979). If it is the latter, the possible time over which non-biological life can exist is vast, though it is unlikely to be eternal (Krauss & Starkman, 2000). The question is not unrelated to curiosity about how far humans can go towards creating artificial intelligence (AI) with sentience and sapience, and what the legacy of our species will be. If we create superintelligent forms of "artificial" *general intelligence* (AGI)—that is if we become as Harari (2017) put it, *homo deus*—we might spawn new non-biological life forms. These life forms might replicate faster than the fastest biological replicators and could conceivably become galactic colonizers. They might also destroy us or make our lives miserable.

Factors that might cause human extinction on much shorter time scales (e.g., in years, decades, or centuries) include (but are not limited to) the intentional or accidental deployment of nuclear, biological, or other weapons, biological threats that evolve 'naturally' without goal-directed technological intervention, climate change (including that caused by bolide collisions), threats from nanotechnology, and technological developments or scientific experiments that turn out to have catastrophic consequences for humanity. Disasters triggered by such "existential risks" (Bostrom, 2002) might fall short of destroying the human species yet cause catastrophic harm, perhaps ending human civilization, as we (or our descendants) know it. Such outcomes would not eradicate humans but could cause massive slumps in human history—slumps that might lock us into dystopian futures or delay technological progress sufficiently long to enable us to encounter an existential risk for which our species is unprepared (Bostrom, 2002; MacAskill, 2022; Ord, 2020).

Therefore, scientific and technological advancement contributes weight to both sides of the species-level (or even life-level) risk/reward equation, affecting our judgments of what precautions might need to be taken. The *precautionary principle* refers to the view that if there is a potentially dangerous threat (e.g., to society or even all of humanity) and its magnitude and evolution are uncertain (e.g., due to incomplete scientific understanding of the problem), then steps to minimize the threat (e.g., through regulation) ought to be to mandated (Sandin, 1999). Human civilizations that adhere to the precautionary principle too strongly could inadvertently preempt the next risky technology that might have been necessary to avoid an impending extinction-level scenario. Excessive doomsterism could generate a self-fulfilling prophecy through risk-averse inaction where, instead, vigorous scientific discovery and technological advancement were needed to survive or perhaps to stave off civilizational decay.

However, if human civilizations adopt technologies that could enable, trigger, or even fully cause an extinction-level sequence of events, it might be only a matter of time until such technologies bring demise to our species. Even infinitesimal probabilities accumulate over vast time scales and there are multiple vulnerability paths to such an outcome (Bostrom, 2019). Given eons, the total probability of disaster to our species due to technological advancement may approach the limit (unity). A fully rational assessment of human choice requires explicit consideration of both the potential benefits and potential harms that would be expected to accrue following the introduction of a new technology. In rational choice theories (Savage, 1954; von Neumann & Morgenstern, 1947), such positive and negative outcomes are to be weighted by their expectations; i.e., the assessed probabilities of their occurrences, whether these are based on relative frequency information, credences, or some combination of these. For technological risks, or other existential or catastrophic risks, whose species-level impacts are extremely uncertain, expected utilities are likely to be calculable with only low (or even no) confidence, placing us in the zone of "radical uncertainty" (Kay & King, 2020).

**Artificial (General) Intelligence**

As with any modern technology, contemporaneous societal and civilizational pressures and scientific breakthroughs continue to shape the evolution of AI. Its effects, much like those of nuclear technology, can and likely will change substantially over time. For instance, the perceived necessity of securing a decisive military advantage over the Axis powers during the Second World War motivated the US's development of nuclear weapons, but this gave way to a military arms race once the US and Soviet Union were locked into the decades-long Cold War. Since then, there has been an expansion of nuclear-armed states and the pressures for expansion continue. While the threat of mutually-assured destruction serves as a potent first-use deterrent, there have been several close calls that raise the question of how long humans can avoid a global nuclear catastrophe (i.e., a nuclear winter) given that there appears to be no harmonious "end of history" in sight (Barrett et al., 2013; Baum, 2015).

Attention to existential risk posed by AI invariably concerns *possible future* AI rather than the current state of AI. Although current AI carries significant risk—for example, algorithmic processes may be unexplainable (Angelov et al., 2021; Vilone & Longo, 2021) or biased and possibly unfair (Fazelpour & Danks, 2021)—and possibly biased for inexplicable reasons (Bostrom & Yudkowsky, 2018)—this is not what most experts regard as *existential* risk from AI. Current AI does not pose an existential threat because its development has not reached the stage where machines have parity with human intelligence across a wide range of tasks and domains. However, many forecasters believe that parity is not far off.

Consider forecasting data from Metaculus, a reputation-based online forecasting and aggregation platform devoted to predicting the scientific and technological outcomes and their consequences (Aguirre, 2016). As of July 20, 2023, the median probability estimate of 1,041 Metaculus forecasters that human-machine parity in intelligence will be reached by 2040 was a staggering 92%.[1] The criterion for scoring parity as having occurred is based on the requirement that the machine system outperform at least two out of three graduate students drawn from elite universities and having backgrounds in physics, mathematics, and computer science, respectively, on a two-hour intelligence test. The machine and three humans could access the Internet but could not consult with other humans. Certainly, more stringent criterion for scoring intelligence parity would be expected to lower the pooled estimate. Nevertheless, other estimates based on takeoff-speed models (e.g., Davidson, 2023) agree that by 2040, AGI will be capable of performing virtually all human tasks regardless of their intelligence requirements. Such models rest on many assumptions, however, and their coherence with forecasters' estimates is just information, not proof.

The prospect of intelligence parity is not *necessarily* bad. AGI that could think as intelligently as humans but at much faster speeds—what Vinge (1993) calls *weak superhumanity* —could accelerate scientific and technological progress that humans might have eventually made but over much longer times. This process has even been given the sticky name, PASTA—Process for Automating Scientific and Technological Advancement (Karnofsky, 2021). If AGI (or PASTA) were necessary to avert an existential calamity (or worse, a mass extinction), having it would be a very good thing. The differential clock speeds of humans and machines, however, casts doubt on the parity notion. As soon as AGI could perform as well on any intelligence test (broadly defined) as the best performing human, it will have exceeded human intelligence since, if nothing else, it will have faster-than-human speed. Unlike humans, AGI would likely be able to access its source code, modify it, and replicate either original or modified copies. These abilities would set off a positive feedback loop constituting what Good (1966) referred to as an *intelligence explosion*. As a result, AGI would evolve very quickly to be *strongly* superintelligent; i.e., it would not only think *faster*, it would think *smarter* than the smartest human (Yukdowsky, 2008).

---

[1]     See https://www.metaculus.com/questions/384/humanmachine-intelligence-parity-by-2040/

It is the prospect of superintelligent AGI that raises concern regarding the continued existence our species and that has caused many AI experts and others alike to call for a pause in advanced AI development (e.g., Future of Life Institute, 2023). If AGI were more powerful than humans, a misalignment of human and AGI goals would likely not end well for humanity. This *alignment problem*—the question of whether AGI will become misaligned with human goals and values—is a threat analysis problem concerning a set of *possible futures* having devastating consequences for humans *as a species*. Since there are no tightly coupled reference classes to draw from in determining the probability of human extinction (or catastrophic demise) from AGI misalignment, estimating existential risk from AGI constitutes an extreme version of the reference class problem (Hájek, 2007; Reichenbach, 1949) with the various pro and con arguments drawing heavily on loose analogies. Therefore, it is of little surprise that estimates of risk vary greatly.

A recent study by Karger et al. (2023) compared the forecasts of 32 AI experts, 48 experts on areas of existential risk other than from AI, and 89 "superforecasters" who demonstrated top-tier forecasting skill on a wide spectrum of short-range topics (Mellers et al., 2015) and who tend to exhibit higher than average coherence across a range of probabilistic judgment tasks (Mellers et al., 2017). The authors found that among various existential threats considered (i.e., AI, nuclear, biological, and climate), existential risk from AI prompted the greatest level of disagreement. The inter-quartile range of estimates of human extinction from AI by 2100, remarkably, was *greater* among AI experts than among other experts from other areas or superforecasters.

As well, Karger et al. (2023) found that the largest absolute disagreement between superforecasters and area-relevant experts (i.e., AI experts for AI topics, nuclear experts for nuclear topics, and so on) focused on AI risk. However, this was largely attributable to the fact that both groups regarded the probability of extinction from AI to be larger than for other risks considered. For instance, whereas superforecasters' median estimated risk of human extinction by 2100 due to AI was 0.38%, the comparable estimated risk due to nuclear technologies was 0.074%. Likewise, whereas AI experts' median estimated risk of extinction by 2100 was 3.0%, the comparable risk from nuclear technologies forecasted by nuclear experts was 0.55%. Remarkably, among both superforecasters and domain experts, the judged probability of human extinction by 2100 due to AI is just over five times higher than the perceived probability of extinction due to nuclear technology. Other existential risks considered by the forecasters in Karger et al.'s study were estimated to be even lower than for nuclear technology, placing AI at the top of the "most probable" list of existential risks among experts and competent generalists (i.e., superforecasters).

**Human Risk Perception**

Humans face a dilemma: for the first time in human history, they appear to be on the brink of developing technologies that might have intelligence superior to their own. Up to now, such examples have been restricted to specific niches, such as games of chess (i.e., Deep Blue beating world champion Garry Kasparov in 1997 in a six-game match) and go (i.e., AlphaGo beating 9 dan Lee Sedol in 2016 and 9 dan world champion Ke Jie in 2017). Such technological feats were historically important, marking the rise in machine competence in arenas once thought to be uniquely human. However, such technologies were not usable by the mainstream of Internet users. But recent access to oracular AI based on large language models that can be applied to a wide range of tasks, notably OpenAI's ChatGPT and GPT4, has made AI accessible to millions of users. For instance, it is estimated that in the first two months since ChatGPT launched, it had reached 100 million monthly active users (Hu, 2023).

These AI oracles have also increased awareness of the prospect of superintelligence and its attendant alignment problem, making these issues much more salient to a broader segment of the public. This is evident in the rapid changes of several AI risk estimates over relatively short timeframes. For instance, the median probability function over time for the Metaculus intelligence parity question noted earlier shows spikes corresponding to the release of these technologies. Since November 2022, when ChatGPT launched, the monthly median Metaculus forecast has increased *monotonically* from 60% to 92%—a 53% increase. Over the same period, the estimated date of an AI catastrophe this century has gotten closer to the present: the aggregate estimate was July 10, 2050, at the start of November 2022 (based on 93 forecasters) and it was August 13, 2038, as of July 20, 2023 (based on 121 forecasters).[2] In a mere ¾ of a year, the estimated time to an AI catastrophe shrank 46% from 10,114 days to 5,504 days. This shift corresponds with changing belief about the time from the first development of weak AGI to superintelligence.[3] At the start of November 2022, the median forecast was 14.24 months and by July 20, 2023, the median was 7.92 months, a 44% reduction.

Other forecaster data from Metaculus indicates that AGI risk is associated with the prospect of the technology arriving too soon to solve the alignment problem. The median probability of solving the alignment problem before weak AGI is developed as judged by 169 Metaculus forecasters on July 20, 2023 was a mere 5%, a 50% reduction from the median 10% estimate given at the start of November 2022 (again when ChatGPT launched).[4]  Likewise, the aggregated conditional probability that if a global catastrophe occurs, it will be caused by AI hovered between 20%-30% from September 2019 (when the number of forecasters reached 100) to September 2022, but has ranged between 30%-50% since then (with 291 forecasters as of July 20, 2023).[5]  In contrast, a comparable conditional probability question focused on nuclear war showed a much flatter curve over the same timeframe, even with media attention to the prospect of nuclear weapons use in the war in Ukraine and the associated dread risk among the public (Scoblic & Mandel, 2022). Although the war in Ukraine continues and, in key respects, represents a proxy war between the US and Russia, the aggregate probability assigned to AI as a cause of world catastrophe is still 24% higher than that assigned to nuclear war (41% vs. 33% [$n = 207$] as of July 20, 2023). Paralleling the results of Karger et al. (2023), comparison to other threat vectors shows even greater gaps in estimated probability of a world catastrophe: 27% ($n = 196$) for a catastrophe caused by biotechnology (e.g., bio-engineered organisms),[6] 5% ($n = 228$) for a catastrophe caused by human-made climate change or geoengineering,[7] and 3% ($n = 160$) for a catastrophe caused a nanotechnology failure.[8]

---

[2] See https://www.metaculus.com/questions/2805/if-there-is-an-artificial-intelligence-catastrophe-this-century-when-will-it-happen/
[3] See https://www.metaculus.com/questions/4123/time-between-weak-agi-and-oracle-asi/

My aim in summarizing these results is not to estimate the actual risk of catastrophe or extinction from AGI (important as that aim is as well), but to highlight a rather stark conclusion about AI risk perception, which is that among the many existential risks humans face, this one appears to be the greatest and on the steepest trajectory of increase at this point in history. This conclusion is only reinforced by turning to the risk perceptions of experts. As noted earlier, Karger et al. (2023) found stronger risk estimates among AI experts than among superforecaster generalists or experts on other risk types.

---

[4] See https://www.metaculus.com/questions/6509/control-problem-solution-before-agi/
[5] See https://www.metaculus.com/questions/1495/ragnar%25C3%25B6k-question-series-if-a-global-catastrophe-occurs-will-it-be-due-to-an-artificial-intelligence-failure-mode/
[6] https://www.metaculus.com/questions/1502/ragnar%25C3%25B6k-question-series-if-a-global-catastrophe-occurs-will-it-be-due-to-biotechnology-or-bioengineered-organisms/
[7] See https://www.metaculus.com/questions/1500/ragnar%25C3%25B6k-question-series-if-a-global-catastrophe-occurs-will-it-be-due-to-either-human-made-climate-change-or-geoengineering/
[8] See https://www.metaculus.com/questions/1501/gc-to-be-caused-by-nanotech-if-it-occurs/

It is not uncommon for the public to dread what experts generally find safe. For instance, the risk of a nuclear disaster has tended to be ranked much higher by the general public than by experts (Slovic et al., 1981), However, this is not the case with AGI. Indeed, Karger et al. (2023) found that among a survey of 912 college graduates, the probability assigned to human extinction from AI by 2100 to be 2%, whereas as noted earlier, AI experts estimated this risk to be 3% with other-domain experts estimating 2%. Clearly, the experts and the public seem to converge in their assessment that AI poses the greatest risk to humans at a global scale. The concern on the part of experts is reinforced by a recent opinion survey of 305 technology innovators, developers, business and policy leaders, researchers and academics recruited by Pew Research Center and Elon University's Imagining the Internet Center (Anderson & Rainee, 2023). The study found that twice as many experts are more concerned than excited (37%) than are more excited than concerned (18%) about the changes in the "humans-plus-tech" evolution they expect to see by 2035, with the modal response (42%) being equally excited and concerned.

**Conclusion**

The significance of congruence between experts and the public over the existential risk posed by AGI is equivocal because the congruence itself does not clarify the accuracy of these judgments. Perhaps experts and the general public are warranted in their risk perceptions or perhaps both groups are overestimating the risk. Both scenarios seem plausible, just as both are consequential. The scientific and technological advantages of AGI (and PASTA) could be the next most important technological development for human civilization. We do not want to thwart our opportunity for prosperity over either the short- or long-term. We must, therefore, assess the cost of applying the precautionary principle too easily. Conversely, we have no adequate methods of ensuring that AGI would be aligned with human goals and values. Modern-day "westerners" find it challenging to relate to the goals, values, and ways of thinking of earlier civilizations, including ones out of which western civilization emerged, such as ancient Greece and Rome. Why would we expect a superintelligent non-biological machine capable of evolving itself at a rate unfathomable by human standards to stay locked into alignment with human interests and values? Such an outcome suggests wishful thinking, if not outright self-deception.

Furthermore, dismissal of the alignment problem puts aside the fact that human interests and values are far from uniform across time and place. Even if *we* could create AGI that was aligned with *our* goals, this might prove to be an inescapable nightmare for future generations of humans who might not share these goals. We must not only anticipate the likelihood of AGI compliance, we must anticipate how even optimal "best-case" scenarios in the present might lock us into intolerable futures, and we must decide whether those are gambles we should be willing to take. We do not have experience writing digital contracts with superintelligences. Clearly, we do not know AGI well enough, but the dread risk associated with the "post-human" era is that we also do not know ourselves well enough. Or perhaps it is also that we know ourselves well enough to know that such powerful technologies are unsafe in human hands—that we are still insufficiently enlightened to handle it.